\documentclass[prl,twocolumn,showpacs,preprintnumbers,amsmath,amssymb,superscriptaddress,unsortedaddress]{revtex4}

\usepackage{graphicx}
\usepackage{dcolumn}
\usepackage{bm}

\begin{document}

\preprint{}

\title{Nonmaximally entangled states can be better for multiple linear optical teleportation}

\author{Joanna Mod{\l}awska}

\affiliation{Faculty of Physics, Adam Mickiewicz University,
Umultowska 85, 61-614 Pozna\'{n}, Poland}

\author{Andrzej Grudka}

\affiliation{Faculty of Physics, Adam Mickiewicz University,
Umultowska 85, 61-614 Pozna\'{n}, Poland}

\affiliation{Institute of Theoretical Physics and Astrophysics,
University of Gda\'{n}sk, 80-952 Gda\'{n}sk, Poland}

\affiliation{National Quantum Information Centre of Gda\'{n}sk, 81-824 Sopot, Poland}

\date{\today}

\begin{abstract}
We investigate multiple linear optical teleportation in the Knill-Laflamme-Milburn scheme with both maximally and nonmaximally entangled states. We show that if the qubit is teleported several times via nonmaximally entangled state then the errors introduced in the previous teleportations can be corrected by the errors introduced in the following teleportations. This effect is so strong that it leads to another interesting phenomenon,  i.e., the total probability of successful multiple linear optical teleportation is higher for nonmaximally entangled states than maximally entangled states.
\end{abstract}

\maketitle

One of the main activities in quantum computation field is linear optical processing of quantum information \cite{Kok}. In particular, the very first experimental demonstration of quantum teleportation was based on linear optics \cite{Bouwmeester}. However, the teleportation
only had a success probability of $25\%$ partially due to the impossibility of performing complete Bell measurement \cite{Lutkenhaus}. In order to perform scalable quantum computation,
it is of essential importance to improve this success probability to
a value close to $100\%$. Recently, Knill, Laflamme, and Milburn (KLM) \cite{Knill}
have shown that the probability of success for the teleportation of
a superposition of vacuum and one photon Fock state can indeed be
increased by using a maximally entangled state of two $N+1$ dimensional
Hilbert spaces encoded in $N$ photons. The probability that
teleportation succeeds is then equal to $1-\frac{1}{N+1}$.
Moreover, when teleportation succeeds, the fidelity of the teleported qubit is equal to $1$. Spedalieri  \emph{et al.} \cite{Spedalieri} generalized their protocol for polarization encoding of a qubit, i.e., when one uses horizontal and vertical polarizations rather than photon number states to represent the logical values $0$ and $1$. Franson \emph{et al.} \cite{Franson} proposed a different scheme,  which  does not require that a qubit has to be teleported with  the perfect fidelity but rather assumes that the qubit is always teleported successfully and aims at maximizing the average fidelity. Their scheme is  based  in fact on the KLM scheme with  another carefully chosen entangled state. Franson \emph{et al.} have shown that their scheme gives better average fidelity of the teleported qubit than the KLM scheme. In \cite{Grudka} we have shown that  if the aim is maximization of the probability of successful teleportation and one requires unit fidelity of the teleported qubit then the state used in the original KLM scheme is optimal. Thus, the maximally entangled state is best suited for single quantum teleportation.
 
In this paper we consider several subsequent linear optical teleportations, i.e., the qubit is teleported from $A$ to $C$, then from $C$ to $D$, and so on, and finally to $B$. We show an interesting phenomenon that when  the final unit fidelity of the teleported qubit  is required after completion  of all teleportations  then the nonmaximally entangled states give higher probability of successful teleportation than the maximally entangled ones. It is surprising because usually maximally entangled states are optimal for information-theoretical tasks \cite{Horodecki}.

Let us begin with description of  a generalization of the KLM scheme of linear optical teleportation to the one which is based on the nonmaximally entangled states \cite{Franson}. In this scheme one uses the following entangled state
\begin{equation}
|t_{N}\rangle=\sum_{i=0}^{N}c_{i}|V\rangle^{i}|H\rangle^{N-i}|H\rangle^{i}|V\rangle^{N-i},
\label{eq:1}
\end{equation}
 where $|V\rangle^{i}$ stands for $|V\rangle_{1}|V\rangle_{2}...|V\rangle_{i}$, i.e., one vertically polarized photon in each of the subsequent modes. Similarly,  $|H\rangle^{N-i}$ stands for $|H\rangle_{i+1}|H\rangle_{i+2}...|H\rangle_{N}$ i.e., one horizontally polarized photon in each of the subsequent modes.  If we use the states $\{|V \rangle^{i}|H \rangle^{N-i}\colon i=0, 1, ..., N\}$ and $\{ |H\rangle^{i}|V \rangle^{N-i}\colon i=0, 1, ..., N\}$ as the orthonormal basis states of Alice's and Bob's Hilbert spaces, respectively, we may  treat the state of Eq.~(1) as  a bipartite entangled state. For $c_{i}=\frac{1}{\sqrt{N+1}}$ we obtain  a maximally entangled state.
In order to teleport a qubit in the state $|\psi\rangle=\alpha|H\rangle+\beta|V\rangle$ Alice applies  the $(N+1)$-point quantum Fourier transform to the input mode and  the $N$ first modes of the state $|t_{N}\rangle$, which is given by: 
\begin{eqnarray}
F_{N} (v_{k}^{\dagger})=\frac{1}{\sqrt{N+1}} \sum_{l_{k}=0}^{N} \omega^{k l_{k}} v_{l_{k}}^{\dagger},\nonumber \\ 
F_{N} (h_{k}^{\dagger})=\frac{1}{\sqrt{N+1}} \sum_{l_{k}=0}^{N} \omega^{k l_{k}} h_{l_{k}}^{\dagger}.
\label{eq:2}
\end{eqnarray}
In the  above equations $v_{k}^{\dagger}$ and $h_{k}^{\dagger}$ are  the creation operators for vertically and horizontally polarized photons   in mode $k$, respectively, and $\omega = e^{i 2 \pi/(N+1)}$.
After this transformation the state of the system is
\begin{widetext}
\begin{equation}
\Big( \frac{1}{\sqrt{N+1}} \Big)^{N+2}\sum_{i=0}^{N} \sum_{l_{0},... l_{N}=0} ^{N} ( \omega^{\sum_{k=0}^{N} k l_{k}} \alpha h_{l_{0}}^{\dagger}v_{l_{1}}^{\dagger}... v_{l_{i}}^{\dagger}h_{l_{i+1}}^{\dagger}...h_{l_{N}}^{\dagger}+\beta v_{l_{0}}^{\dagger}v_{l_{1}}^{\dagger}... v_{l_{i}}^{\dagger}h_{l_{i+1}}^{\dagger}...h_{l_{N}}^{\dagger} ) c_{i} |VAC\rangle_{0...N} |H\rangle^{i} |V\rangle^{N-i} .
\end{equation}
\end{widetext}
Note that in the first term there  are $i$ creation operators for vertically polarized photons while in the second term there  are $i+1$ creation operators for vertically polarized photons. Next, Alice measures the total  number of vertically polarized photons and horizontally polarized photons in each of the first $N+1$ modes. If she  detects $v_{j}$ vertically polarized photons  and $h_{j}$ horizontally polarized photons in mode $j$ then the state of the last $N$ modes is 
\begin{widetext}
\begin{equation}
|H\rangle^{m-1}\frac{1}{\sqrt{p(m)}}(\alpha c_{m}|H\rangle+\beta c_{m-1}\omega^{-\sum_{j=0}^{N} j (v_{j}+h_{j})}|V\rangle)|V\rangle^{N-m},
\end{equation}
\end{widetext}
where $0<m=\sum_{j=0}^{N}v_{j}<N+1$ is the total number of  vertically polarized photons  detected and $p(m)=|\alpha c_{m}|^2+|\beta c_{m-1}|^2$ is the total probability of  detecting $m$ vertically polarized photons. However, if Alice  detects $0$ or $N+1$ vertically polarized photons, then the state of the qubit is irreversibly destroyed, which happens with  the average probability  of $\frac{1}{2}(|c_{0}|^2+ |c_{N}|^2)$. One can see that the modified state of the teleported qubit is found in the $N+m$th mode.  After correction of the phase, this state becomes
\begin{equation}
|\psi_{m}\rangle=\frac{1}{\sqrt{p(m)}}(\alpha c_{m}|H\rangle+\beta c_{m-1}|V\rangle).
\label{eq:3}
\end{equation}
The qubit can be returned to its original state by performing  the generalized measurement given by  the Kraus operators:
\begin{eqnarray}
E_{S}=\frac{c_{m-1}}{c_{m}}|H\rangle \langle H|+|V\rangle \langle V|,\nonumber\\
E_{F}=\sqrt{1-\left|\frac{c_{m-1}}{c_{m}}\right|^2}|H\rangle \langle H|,
\label{eq:5}
\end{eqnarray}
for $|c_{m-1}| \leq |c_{m}|$. A similar measurement exists if $|c_{m-1}| > |c_{m}|$.
When $E_{S}$ is applied, then the qubit ends in its original state $|\Psi \rangle = \alpha |H \rangle + \beta |V \rangle$. The probability of successful error correction is:
\begin{equation}
p(S|m)=\langle\psi_{m}|E_{S}^{\dagger}E_{S}|\psi_{m}\rangle=\frac{|c_{m-1}|^2}{p(m)}.
\label{eq:7}
\end{equation}
In \cite{Grudka} we described how such a measurement can be implemented experimentally with linear optics. In order to obtain the total probability of  a successful teleportation, we have to sum up the joint probabilities of  detecting $m$ vertically polarized photons and  a successful error correction.  Let us recall that if $0$ or $N+1$ photons are detected, then the teleportation fails. Hence, we restrict  the summation over $m$ from $1$ to $N$. We obtain:
\begin{eqnarray}
p(S)=\sum_{m=1}^{N}p(S,m)=
\nonumber\\
=\sum_{m=1}^{N}p(S|m)p(m)=\sum_{m=1}^{N}\text{min}\{|c_{m-1}|^2, |c_{m}|^2\}.
\label{eq:9}
\end{eqnarray}

\begin{figure}
\includegraphics [width=9truecm]{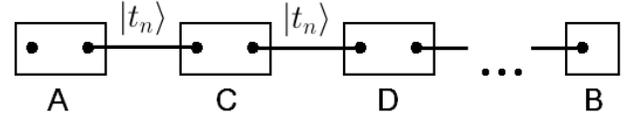}
\caption{\label{fig:1} Several subsequent teleportations.  A better strategy is  not to perform error correction at $C$, $D$ and so on  but to perform it at $B$ after completion of all teleportations.}
\end{figure}

Let us now suppose that the qubit is to be teleported once again (see Fig.~\ref{fig:1}). Then the simplest strategy is to perform the first teleportation  followed by the error correction, then the second teleportation  followed by the error correction. However, it is not  the optimal strategy. Let us, thus, assume that we do not correct the error introduced in the first teleportation and teleport the qubit once again with the use of  the identical entangled state. After the second teleportation, the state of the qubit is
\begin{equation}
|\psi_{m,n}\rangle=\frac{1}{\sqrt{p(m,n)}}(\alpha c_{m} c_{n}|H\rangle+\beta c_{m-1} c_{n-1}|V \rangle),
\label{eq:10}
\end{equation}
where $p(m,n)$ is the joint probability of detecting $m$ vertically polarized photons in the first teleportation and $n$ vertically polarized photons in the second teleportation, and is  given by: 
\begin{equation}
p(m,n)=p(n|m)p(m)=|\alpha c_{m} c_{n}|^2+|\beta c_{m-1} c_{n-1}|^2.
\label{eq:11}
\end{equation}
If $c_{m}=c_{n-1}$ and $c_{n}=c_{m-1}$, then the state of the qubit is
\begin{equation}
|\psi_{m,n}\rangle=\alpha |H\rangle+\beta |V\rangle;
\label{eq:12}
\end{equation}
i.e., it is the original state of the qubit and we do not have to perform  the error correction. The second teleportation corrected the error introduced by the first teleportation. We call this effect  the error self-correction. A similar effect occurs for entanglement swapping as considered by Acin, Cirac, and Lewenstein \cite{Acin} (see also:  \cite{Perseguers1, Bose1, Bose2, Hardy3}). In general, if $|c_{m-1} c_{n-1}| <|c_{m} c_{n}|$, then one can recover the original state of the qubit by performing generalized measurement given by  the Kraus operators:
\begin{eqnarray}
E_{S}=\frac{c_{m-1} c_{n-1}}{c_{m} c_{n}}|H\rangle \langle H|+|V \rangle \langle V|,\nonumber\\
E_{F}=\sqrt{1-\left|\frac{c_{m-1} c_{n-1}}{c_{m} c_{n}}\right|^2}|H \rangle \langle H|.
\label{eq:13}
\end{eqnarray}
A similar measurement exists if $|c_{m-1} c_{n-1}| > |c_{m} c_{n}|$. The joint probability of detecting $m$ vertically polarized photons in the first teleportation and $n$ vertically polarized photons in the second teleportation and  the successful error correction is
\begin{equation}
p(S,m,n)=\text{min}(|c_{m} c_{n}|^2, |c_{m-1} c_{n-1}|^2).
\label{eq:14}
\end{equation}
On the other hand, if we performed the first teleportation followed by the error correction and the second teleportation followed by the error correction, then the probability of detecting $m$ vertically polarized photons in the first teleportation and $n$ vertically polarized photons in the second teleportation and  the successful  correction of both errors would be
\begin{equation}
p'(S,m,n)=\text{min}(|c_{m}|^2, |c_{m-1}|^2) \text{min}(|c_{n}|^2,|c_{n-1}|^2),
\label{eq:15}
\end{equation}
which is lower or equal to the previous probability. Moreover, if $(|c_{m-1}|  > |c_{m}|$ and $|c_{n-1}|  < |c_{n}|)$ or $(|c_{m-1}|  < |c_{m}|$ and $|c_{n-1}|  >|c_{n}|)$, then the probability $p'(S,m,n)$ is  lower than the probability $p(S,m,n)$. We conclude that it is better to perform  the error correction at the end when all teleportations  were completed.

Let us now suppose that we perform $M$ subsequent teleportations with the use of  the identical entangled states of Eq.~(\ref{eq:1}). A straightforward calculation gives the following probability of  detecting $m_{1}$,  $m_{2}$,..., $m_{M}$ vertically polarized photons in the first, second,..., $M$th  teleportation and  the final successful error correction
\begin{eqnarray}
p(S, m_{1}, m_{2},...m_{M})=
\nonumber\\
\text{min}(|c_{m_{1}} c_{m_{2}}...c_{m_{M}}|^2, |c_{m_{1}-1} c_{m_{2}-1}... c_{m_{M}-1}|^2).
\label{eq:16}
\end{eqnarray}
In order to obtain the total probability of successful multiple teleportation we have to sum these probabilities over $m_{1}$,  $m_{2}$,...,  $m_{M}$ ranging from $1$ to $N$. We obtain
\begin{eqnarray}
p(S)=\sum_{ m_{1}=1, m_{2}=1, ..., m_{M}=1}^{N}
\nonumber\\
\text{min}(|c_{m_{1}} c_{m_{2}}...c_{m_{M}}|^2, |c_{m_{1}-1} c_{m_{2}-1}... c_{m_{M}-1}|^2).
\label{eq:17}
\end{eqnarray}

\begin{figure}
\includegraphics {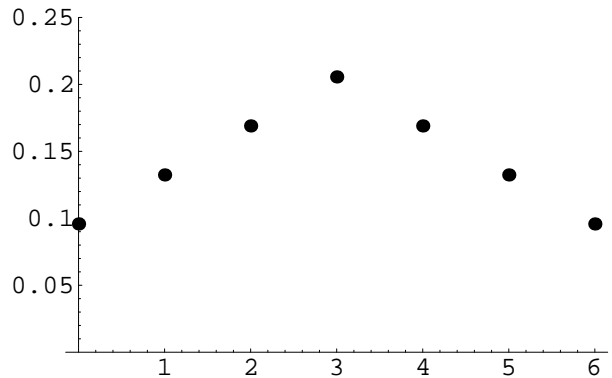}
\caption{\label{fig:2} Exemplary series of coefficients $|c_{i}|^2$ of entangled state of  Eq.~(\ref{eq:18})  ($x=0.0366$).}
\end{figure}

Let us now take the following six-photon entangled state whose coefficients  $c_{i}$  depend on the parameter $x$ (see Fig.~\ref{fig:2})
\begin{eqnarray}
|t_{6}\rangle= \sum_{i=0}^{6} \sqrt{\frac{1-9x}{7}+ (3-|i-3| )x}
\nonumber\\
|V\rangle^{i}|H\rangle^{6-i}|H\rangle^{i}|V\rangle^{6-i}.
\label{eq:18}
\end{eqnarray}
The coefficients are symmetric around $i=3$  and the parameter $x$ is the slope of the line connecting  the points $(i,|c_{i}|^2)$. For $x=0$ the state is maximally entangled. Note that for $x>0$, the smallest coefficients are $c_{0}$ and $c_{6}$. On the other hand, the average probability that the state of the teleported qubit will be irreversibly destroyed during teleportation (and before error correction) is $\frac{1}{2}(|c_{0}|^2+ |c_{6}|^2)$. We can lower this probability by lowering the coefficients $c_{0}$ and $c_{6}$. We should  remember that in such a case we increase the probability that the state will be irreversibly destroyed during error correction.  Let us calculate with the help of Eq.~(\ref{eq:17}) the total probability of successful six subsequent teleportations of a qubit with  the error correction at the end. 
\begin{figure}
\includegraphics{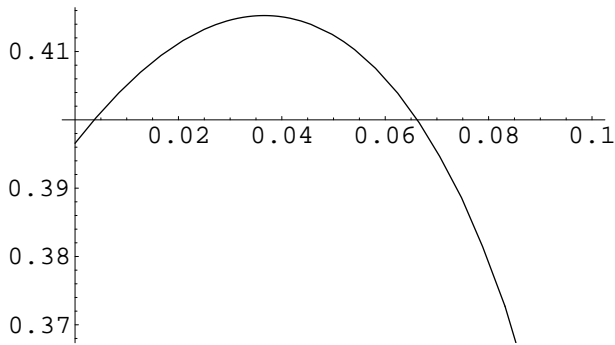}
\caption{\label{fig:3} Probability of successful six teleportations  with entangled state of  Eq.~(\ref{eq:18})  as a function of $x$.  For $x=0$ the state is maximally entangled while for $x \neq 0$ it is nonmaximally entangled. }
\end{figure}

In Fig.~\ref{fig:3} we present how probability of successful
multiple teleportation depends on the parameter $x$. For $x=0$, the probability of successful teleportation  is $p=0.3965$. However,
for $x \geq 0$ this probability slowly increases  with $x$ reaching its
maximal value $p=0.4152$ for $x=0.0366$. Hence, we obtain an
interesting phenomenon -- the probability of  the successful multiple
teleportation is greater for nonmaximally entangled states than  for 
maximally entangled ones. It should be compared with  the probability of
successful single teleportation which reaches always its maximal
value for maximally entangled state \cite{Grudka}.  Let us also
 point out that the probability for $x=0$ is equal to the product of
probabilities of each successful teleportation, i.e.,
$p=(\frac{6}{7})^{6}=0.3965$.  We can see that if we use nonmaximally entangled states we have  an increase in the probability $p(0.0366)-p(0)=0.0187$.   The relative increase in
the probability is $\frac{p(0.0366)-p(0)}{p(0)}=0.0471$. Thus, the
use of nonmaximally entangled states contributes in about $5 \%$ to
the total probability. The effect may be even stronger when one
increases the number of photons in the entangled state and/or the
number of subsequent teleportations. It is also interesting to
calculate the probability $p'(0.0366)$ of successful teleportation
when one performs error correction between subsequent
teleportations. We obtain $p'(0.0366)=0.2511$. This probability is
 lower than the probability of successful teleportation with  the final
error correction  $p(0.0366)=0.4152$. The increase in the probability due
to  the error self-correction is $p(0.0366)-p'(0.0366)=0.1641$ and the
relative increase is
$\frac{p(0.0366)-p'(0.0366)}{p'(0.0366)}=0.6535$.

 Let us now have a look at the origin of this effect. The teleportation does not succeed when one of the senders  detects $0$ or $N+1$ vertically polarized photons which happens with  the average probabilities  of $\frac{1}{2}|c_{0}|^2$ and  $\frac{1}{2}|c_{N}|^2$, respectively. We can decrease this probabilities by decreasing the coefficients $c_{0}$ and $c_{N}$. If we do it and perform single teleportation, then  the error correction is needed. The probabilistic nature of  the error correction decreases the total probability of successful teleportation and there is no gain. However, if we perform several teleportations with no error correction between subsequent teleportations, then  the error self-correction may occur. This error self-correction may correct  the errors or increase the probability of  a successful error correction at the end. This effect allows us to obtain higher probability of successful teleportations with nonmaximally entangled states which have smoothly lowered the probabilities of having $0$ and $N$ vertically polarized photons. 

In summary, we have considered several subsequent teleportations of a qubit in the KLM scheme. We have shown how the errors introduced in the previous teleportations can be corrected by the errors introduced in the following teleportations. This effect leads to  an interesting new phenomenon. Namely nonmaximally entangled states can be better for multiple linear optical teleportation. This strange behavior is connected to the fact that with linear optics one cannot perform  the complete Bell measurement and hence, quantum teleportation can be implemented only probabilistically \cite{Lutkenhaus}. We believe that our research will lead to deeper understanding of manipulation of entanglement with local linear optical operations and classical communication which are a natural analog of local operations and classical communication usually considered in entanglement theory. Our result may have applications in linear optical quantum computation and quantum networks.

\begin{acknowledgments} We thank Adam Miranowicz for independent numerical checking of our results. We thank Oskar Baksalary, Ryszard Horodecki, and our Referees for comments on the manuscript. One of the authors (A.G.) was supported by the State Committee for Scientific Research Grant No. 1 P03B 014 30 and by the European Commission through the Integrated Project FET/QIPC ÒSCALAÓ.
\end{acknowledgments}

\end{document}